\documentclass[superscriptaddress,aip,amsmath,amssymb,floatfix,reprint,raggedbottom]{revtex4-1}
\usepackage[pdftex]{graphicx}
\usepackage{amsmath}
\usepackage{lipsum}
\usepackage{color}
\usepackage{soul}
\usepackage{balance}
\usepackage{subfigure}
\usepackage{fancyhdr}
\usepackage{braket}
\usepackage{threeparttable}
\usepackage{booktabs}
\usepackage{enumitem}
\usepackage{listings}

\fancypagestyle{firststyle}
{
\fancyhf{}

\rfoot{\thepage}
\pagestyle{fancy}
}

\usepackage{fancyhdr}

\begin{document}
\title{Probabilistic Circuits for Autonomous Learning: A simulation study}
\author{Jan Kaiser}
\email{kaiser32@purdue.edu}
\author{Rafatul Faria}
\author{Kerem Y. Camsari}
\author{Supriyo Datta}
\affiliation{Department of Electrical and Computer Engineering, Purdue University, West Lafayette, IN, 47906 USA}
\date{\today}

\begin{abstract}
Modern machine learning is based on powerful algorithms running on digital computing platforms and there is great interest in accelerating the learning process and making it more energy efficient. In this paper we present a fully autonomous probabilistic circuit for fast and efficient learning that makes no use of digital computing. Specifically we use SPICE simulations to demonstrate a clockless autonomous circuit where the required synaptic weights are read out in the form of analog voltages. This allows us to demonstrate a circuit that can be built with existing technology to emulate the Boltzmann machine learning algorithm based on gradient optimization of the maximum likelihood function. Such autonomous circuits could be particularly of interest as standalone learning devices in the context of mobile and edge computing.
\end{abstract}

\maketitle
\thispagestyle{firststyle}

\section{Introduction}

Machine learning, inference and many other emerging applications\cite{schuman_survey_2017} make use of stochastic neural networks comprising (1) a binary stochastic neuron (BSN)\cite{ackley_learning_1985,neal_connectionist_1992} and (2) a synapse that constructs the inputs $I_i$ to the $i^{th}$ BSN from the outputs $m_j$ of all other BSNs.

The output $m_i$ of the $i^{th}$ BSN fluctuates between +1 and -1 with a probability controlled by its input
\begin{equation}
m_i (t+\tau_{N}) = {\rm{sgn}}\left[ \tanh \left( {I_i(t)}\right) - r\right] 
\label{eq:binary_stochastic_neuron}
\end{equation}
\noindent where $r$ represents a random number in the range $\left[-1,+1\right]$, and $\tau_{N}$ is the time it takes for a neuron to provide a stochastic output $m_i$ in accordance with a new input $I_i$.

Usually the synaptic function, $ I_i (\{m\})$ is linear and is defined by a set of weights $W_{ij}$ such that
\begin{equation}
I_i (t+\tau_S) =  \sum_{j} W_{ij} m_j(t) 
\label{eq:synaptic_function}
\end{equation}
\noindent where $\tau_{S}$ is the time it takes to recompute the inputs $\{I\}$ everytime the outputs $\{m\}$ change. Typically Eqs.\eqref{eq:binary_stochastic_neuron},\eqref{eq:synaptic_function} are implemented in software, often with special accelerators for the synaptic function using GPU/TPUs.\cite{jouppi_google_2016,schmidhuber_deep_2015}

The time constants  $\tau_N$ and $ \tau_S$ are not important when Eqs.\eqref{eq:binary_stochastic_neuron} and \eqref{eq:synaptic_function} are implemented on a digital computer using a clock to ensure that neurons are updated sequentially and the synapse is updated  between any two updates. But they play an important role in clockless operation of autonomous hardware that makes use of the natural physics of specific systems to implement Eqs.\eqref{eq:binary_stochastic_neuron} and\eqref{eq:synaptic_function} approximately. A key advantage of using BSNs is that Eq. \eqref{eq:binary_stochastic_neuron} can be implemented compactly using stochastic magnetic tunnel junctions (MTJs) as shown in Camsari et al. \cite{camsari_stochastic_2017,camsari_implementing_2017}, while resistive or capacitive crossbars can implement Eq. \eqref{eq:synaptic_function}.\cite{hassan_voltage-driven_2019} It has been shown that such hardware implementations can operate autonomously without clocks, if \textit{the BSN operates slower than the synapse}, that is, if $\tau_N >> \tau_S$ shown by Sutton et al. \cite{sutton_autonomous_2019}

Stochastic neural networks defined by Eqs. \eqref{eq:binary_stochastic_neuron} and \eqref{eq:synaptic_function} can be used for inference whereby the weights $W_{ij}$ are designed such that the system has a very high probability of visiting configurations defined by $\{m\}= \{v\}_n$, where $\{v\}_n$ represents a specified set of patterns. However, the most challenging and time-consuming part of implementing a neural network is not the inference function, but the learning required to determine the correct weights $W_{ij}$  for a given application. This is commonly done using powerful cloud-based processors and there is great interest in accelerating the learning process and making it more energy efficient so that it can become a routine part of mobile and edge computing.
 
In this paper we present a new approach to the problem of fast and efficient learning that makes no use of digital computing at all. Instead it makes use of the natural physics of a fully autonomous probabilistic circuit composed of standard electronic components like resistors, capacitors and transistors along with stochastic MTJs. 
\begin{figure*}[!t]
    \setlength\abovecaptionskip{-0.5\baselineskip}
    \centering
    \includegraphics[width=1\linewidth]{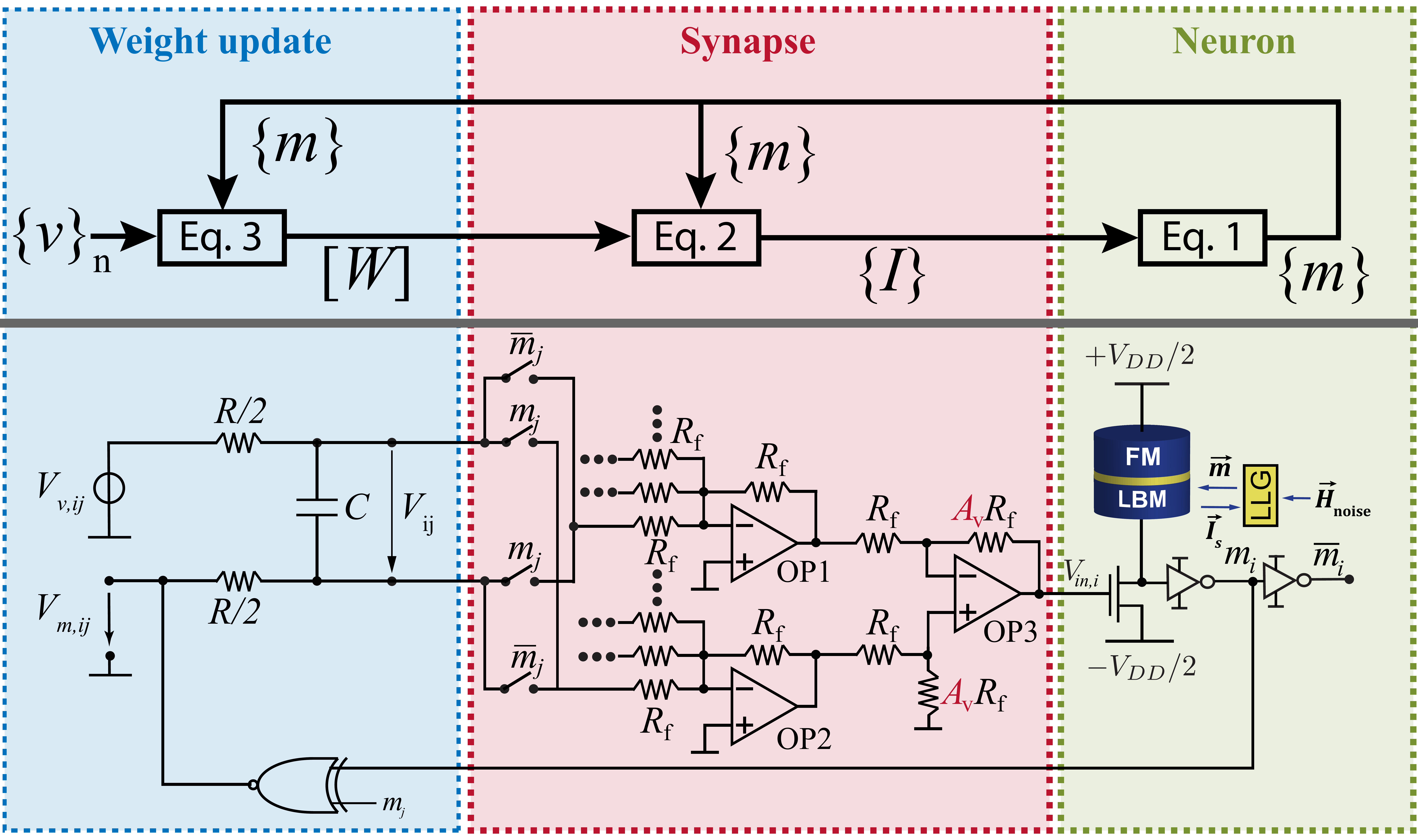} 
    \vspace{0.1in}
    \caption{\textbf{Clockless learning circuit} designed to emulate Eqs. (1)-(3) autonomously.}
    \label{fig:autonomous_learning_circuit}
\end{figure*}

We focus on a fully visible Boltzmann machine (FVBM), a form of stochastic recurrent neural network, for which the most common learning algorithm is based on the gradient ascent approach to optimize the maximum likelihood function.\cite{koller_probabilistic_2009,carreira-perpinan_contrastive_2005} We use a slightly simplified version of this approach, whereby the weights are changed incrementally according to
\begin{equation*}
W_{ij}(t+\Delta t)  = W_{ij}(t) + \epsilon [v_i v_j - m_i m_j - \lambda W_{ij}(t)]
\end{equation*}
\noindent where $\epsilon$ is the learning parameter and $\lambda$ is the regularization parameter.\cite{ng_feature_2004} The term $v_iv_j$ is the correlation between the $i$th and the $j$th entry of the training vector $\{v\}_n$. The term $m_im_j$ corresponds to the sampled correlation taken from the model's distribution. The advantage of this network topology is that the learning rule is local since it requires only information of the two neurons $i$  and $j$ connected by weight $W_{ij}$. In addition, the learning rule can tolerate stochasticity for example in the form of sampling noise which makes it an attractive algorithm to use for hardware machine learning.\cite{carreira-perpinan_contrastive_2005,fischer_training_2014,ernoult_using_2019}

For our autonomous operation we replace the equation above with its continuous time version ($\tau_{L}$: learning time constant)
\begin{equation}
\frac{dW_{ij}}{dt} = \frac{v_i v_j - m_i m_j - \lambda W_{ij}}{\tau_{L}}
\label{eq:learning_rule}
\end{equation}
\noindent which we translate into an RC circuit by associating $W_{ij}$ with the voltage on a capacitor C driven by a voltage source $(V_{v,ij} - V_{m,ij})$ with a series resistance R (Fig.\eqref {fig:autonomous_learning_circuit}):
\begin{equation}
C \frac{dV_{ij}}{dt} = \frac{V_{v,ij} - V_{m,ij} -V_{ij}}{R} 
\label{eq:learning_circuit}
\end{equation}
\noindent with $v_iv_j=V_{v,ij}/(V_{DD}/2)$ and $m_im_j=V_{m,ij}/(V_{DD}/2)$. From Fig. \ref{fig:autonomous_learning_circuit} and comparing Eqs. \eqref{eq:learning_rule},\eqref{eq:learning_circuit} it is easy to see how the weights and the learning and regularization parameters are mapped into circuit elements: $W_{ij}=A_v V_{ij}/V_0$, $\lambda = V_0/(A_v V_{DD}/2)$ and $\tau_{L}=\lambda RC$ where $A_v$ is the voltage gain of OP3 in Fig. \ref{fig:autonomous_learning_circuit} and $V_0$ is the reference voltage of the BSN. For proper operation the learning time scale $\tau_L$ has to be much larger than the neuron time $\tau_N$ to be able to collect enough statistics throughout the learning process.

A key element of this approach is the representation of the weights $W$ with $voltages$ rather than with programmable $resistances$ for which memristors and other technologies are still in development.\cite{li_review_2018} By contrast the charging of capacitors is a textbook phenomenon, allowing us to design a learning circuit that can be built today with established technology. The idea of using capacitor voltages to represent weights in neural networks has been presented by several authors for different network topologies in analog learning circuits.\cite{kim_analog_2017,schneider_analog_1993,card_learning_1994,sung_perspective:_2018} The use of capacitors has the advantage of having a high level of linearity and symmetry for the weight updates during the training process.\cite{li_capacitor-based_2018}

In section \ref{sec: methods}, we will describe such a learning circuit that emulates Eqs. \eqref{eq:binary_stochastic_neuron}-\eqref{eq:learning_rule}. The training images or patterns $\{v\}_{n}$  are fed in as electrical signals into the input terminals, and the synaptic weights $W _{ij}$ can then be read out in the form of voltages from the output terminals. Alternatively the values can be stored in a non-volatile memory from which they can subsequently be read and used for inference. In section \ref{sec: results}, we will present SPICE simulations demonstrating the operation of this autonomous learning circuit.

\section{Methods}
\label{sec: methods}

The autonomous learning circuit has 3 parts where each part represents one of the three Eqs. \eqref{eq:binary_stochastic_neuron}-\eqref{eq:learning_rule}. On the left hand side of Fig. \ref{fig:autonomous_learning_circuit}, the training data is fed into the circuit by supplying a voltage $V_{v,ij}$ which is given by the $i$th entry of the bipolar training vector $v_i$ multiplied by the $j$th entry of the training vector $v_j$ and scaled by the supply voltage $V_{DD}/2$. The training vectors can be fed in sequentially or as an average of all training vectors. The weight voltage $V_{ij}$ across capacitor $C$ follows Eq. \eqref{eq:learning_circuit} where $V_{v,ij}$ is compared to voltage $V_{m,ij}$ which represents correlation of the outputs of BSNs $m_i$ and $m_j$. Voltage $V_{m,ij}$ is computed in the circuit by using an XNOR gate that is connected to the output of BSN $i$ and BSN $j$. 
The synapse in the center of the circuit connects weight voltages to neurons according to Eq. \eqref{eq:synaptic_function}. Voltage $V_{ij}$ has to be multiplied by 1 or -1 depending on the current value of $m_j$. This is accomplished by using a switch which connects either the positive or the negative node of $V_{ij}$ to the operational amplifiers OP1 and OP2. Here, OP1 accumulates all negative contributions and OP2 accumulates all positive contributions of the synaptic function. The differential amplifier OP3 takes the difference between the output voltages of OP2 and OP1 and amplifies the voltage by amplification factor $A_v$. This voltage conversion is used to control the voltage level of $V_{ij}$ in relation to the input voltage of each BSN. The voltage level at the input of the BSN is fixed by the reference voltage of the BSN which is $V_0$. However, the voltage level of $V_{ij}$ can be adjusted and utilized to adjust the regularization parameter $\lambda$ in the learning rule (Eq. \eqref{eq:learning_rule}). 
The functionality of the BSN is described by Eq. \eqref{eq:binary_stochastic_neuron} where the dimensionless input is given by $I_i(t)=V_{i,in}(t)/V_0$. This relates the voltage $V_{ij}$ to the dimensionless weight by $W_{ij}=A_v V_{ij}/V_0$. The hardware implementation of the BSN uses a stochastic MTJ in series with a transistor as presented by Camsari et al. \cite{camsari_implementing_2017} Due to thermal fluctuations of the low-barrier magnet (LBM) of the MTJ the output voltage of the MTJ fluctuates randomly but with the right statistics given by Eq. \ref{eq:binary_stochastic_neuron}. The time dynamics of the LBM can be obtained by solving the stochastic Landau-Lifshitz-Gilbert (LLG) equation. Due to the fast thermal fluctuations of the LBM in the MTJ, Eq. \eqref{eq:binary_stochastic_neuron} can be evaluated on a subnanosecond timescale leading to fast generation of samples. \cite{kaiser_subnanosecond_2019,hassan_low-barrier_2019}

Fig. \ref{fig:autonomous_learning_circuit} just shows the hardware implementation of one weight and one BSN. The size of the whole circuit depends on the size of the training vector $N$. For every entry of the training vector one BSN is needed. The number of weights which is the number of RC-circuits is given by $N(N-1)/2$ where every connection between BSNs is assumed to be reciprocal. To learn biases another $N$ RC-circuits are needed.

The learning process is captured by Eqs. \eqref{eq:learning_rule} and \eqref{eq:learning_circuit}. The whole learning process has similarity with the software implementation of persistent contrastive divergence (PCD)\cite{tieleman_training_2008} since the circuit takes samples from the model's  distribution ($V_{m,ij}$) and compares it to the target distribution ($V_{v,ij}$) without reinitializing the Markov Chain after a weight update. During the learning process voltage $V_{ij}$ reaches a constant average value where $\frac{dV_{ij}}{dt} \approx 0$. This voltage $V_{ij}=V_{ij,\mathrm{learned}}$ corresponds to the learned weight. 

For inference the capacitor $C$ is replaced by a voltage source of voltage $V_{ij,\mathrm{learned}}$. Consequently, the autonomous circuit will compute the desired functionality given by the training vectors. In general, training and inference have to be performed on identical hardware in order to learn around variations (see supplemental material for more details).
It is important to note that in inference mode this circuit can be used for optimization by performing electrical annealing. This is done by increasing all weight voltages $V_{ij}$ by the same factor over time. In this way the ground state of a Hamiltonian like the Ising Hamiltionian can be found.\cite{sutton_intrinsic_2017,camsari_scalable_2019}
\begin{figure}[!t]
    \setlength\abovecaptionskip{-0.5\baselineskip}
    \centering
    \includegraphics[width=1\linewidth]{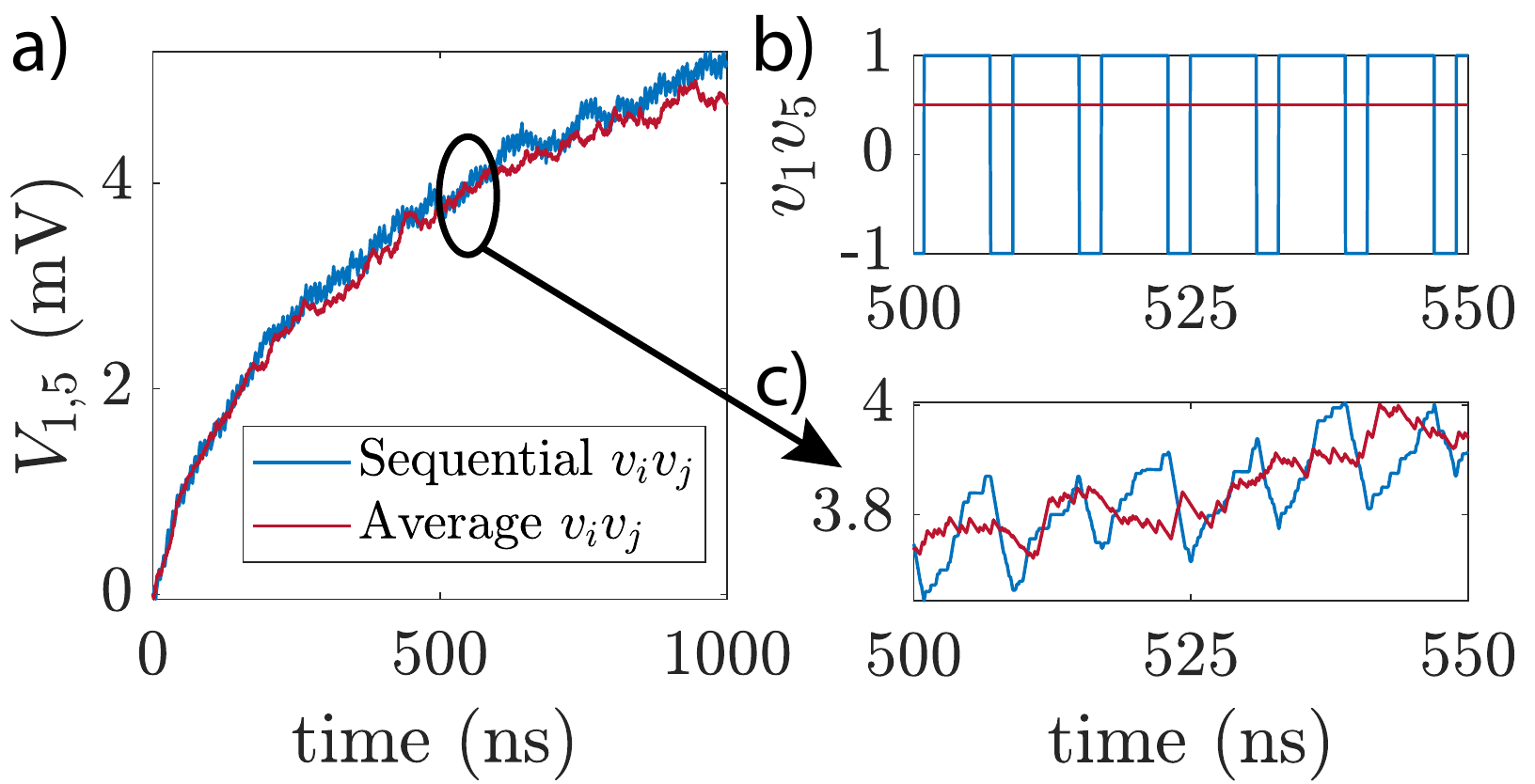} 
    \vspace{0.1in}
    \caption{\textbf{Feeding of training data into the circuit} a) Weight voltage $V_{1,5}$ over time for sequential and average feeding in of the correlation between visible unit $i$ and visible unit $j$ for training a full adder. b) Correlation $v_1v_5$ vs. time $t$. All eight lines of the truth table of a full adder are cycled through where every vector is shown for time $T=1$ ns at a time. c)  Enlarged version of subfigure a). For sequential feeding in of data, the voltage change in $v_1v_5$ directly affects $V_{1,5}$.}
    \label{fig:fig_vivj}
\end{figure}

\begin{center}
\begin{table}
 \begin{tabular}{|c | c | c | c | c | c | c|} 
 \hline
\begin{tabular}{@{}c@{}} $\mathbf{A}$ \\ $v_1$\end{tabular} & \begin{tabular}{@{}c@{}} $\mathbf{B}$ \\ $v_2$\end{tabular} & \begin{tabular}{@{}c@{}} $\mathbf{C_\mathrm{in}}$ \\ $v_3$\end{tabular} & \begin{tabular}{@{}c@{}} $\mathbf{S}$ \\ $v_4$\end{tabular} & \begin{tabular}{@{}c@{}} $\mathbf{C_\mathrm{out}}$ \\ $v_5$\end{tabular}  & \textbf{Dec} & $P_\mathrm{Ideal}$\\ 
 \hline\hline
 -1 & -1 & -1 & -1 & -1 & 0 & 0.125 \\ 
 \hline
 -1 & -1 & 1 & 1 & -1 & 6 & 0.125 \\ 
 \hline
 -1 & 1 & -1 & 1 & -1 & 10 & 0.125 \\ 
 \hline
 -1 & 1 & 1 & -1 & 1 & 13 & 0.125 \\ 
 \hline
 1 & -1 & -1 & 1 & -1 & 18& 0.125\\ 
  \hline
 1 & -1 & 1 & -1 & 1 & 21 & 0.125 \\ 
 \hline  
 1 & 1 & -1 & -1 & 1 & 25 & 0.125 \\ 
 \hline 
 1 & 1 & 1 & 1 & 1 & 31 & 0.125 \\ 
 \hline
\end{tabular}
 \caption{\textbf{Truth Table of a full adder} Every 0 in the binary representation of the full adder is replaced by -1 in the bipolar representation. "Dec" represents the decimal conversion of each line. $P_\mathrm{Ideal}$ is the ideal probability distribution were every line's probability is $p=1/8=0.125$.}
\label{table: FA}
\end{table}
\end{center}
\section{Results}
\label{sec: results}
In this section the autonomous learning circuit in Fig. \ref{fig:autonomous_learning_circuit} is simulated in SPICE. We show how the proposed circuit can be used for both inference and learning. As examples, we demonstrate the learning on a full adder and on 5x3 digit images. The BSN models are simulated in the framework developed by Camsari et al.\cite{camsari_modular_2015} For all SPICE simulations the following parameters are used for the stochastic MTJ in the BSN implementation: Saturation magnetization $M_S=1100 \ \mathrm{emu/cc}$, LBM diameter $D=22$ nm, LBM thickness $l=2$ nm, TMR=110\%, damping coefficient $\alpha=0.01$, temperature $T=300 \ \mathrm{K}$ and demagnetization field $H_D=4 \pi M_S$ with $V=(D/2)^2 \pi l$. For the transistors, 14 nm HP-FinFET Predictive Technology Models (PTM)\footnote{http://ptm.asu.edu/} are used with fin number $fin=1$ for the inverters and $fin=2$ for XNOR-gates. Ideal operational amplifiers and switches are used in the synapse. The characteristic time of the BSNs $\tau_N$ is in the order of 100 ps\cite{hassan_low-barrier_2019} and much larger than the time it takes for the synaptic connections, namely the resistors and operational amplifiers, to propagate BSN outputs to neighboring inputs. It has to be noted that in principle other hardware implementations of the synapse for computing Eqn. \eqref{eq:synaptic_function} could be utilized as long as the condition $\tau_N \gg \tau_S$ is satisfied.\\
\subsection*{Learning addition}
As first training example, we use the probability distribution of a full adder. The FA has 5 nodes and 10 weights that have to be learned. In the case of the FA training, no biases are needed. The probability distribution of a full adder with bipolar variables is shown in table \ref{table: FA}. To learn this distribution the correlation terms $v_iv_j$ in the learning rule have to be fed into the voltage node $V_{v,ij}$. The correlation is dependent on what training vector / truth table line is fed in. For the second line of the truth table for example $v_1v_2=-1 \cdot -1 =1$ and $v_1v_3=-1 \cdot 1 =-1$ with A being the first node, B the second  node and so on. In Fig.  \ref{fig:fig_vivj}  b) the correlation $v_1v_5$ is shown. For the sequential case the value of $v_1v_5$ is obtained by circling through all lines of the truth table where each training vector is shown for $1$ ns. $A$ and $C_\mathrm{out}$ in table \ref{table: FA} only differ in the fourth and fifth line for which $v_1v_5=-1$. For all other cases $v_1v_5=1$. The average of all lines is shown as red solid line. Fig. \ref{fig:fig_vivj} a) shows the weight voltage $V_{ij}$ with $i=1$ and $j=5$ for FA learning and the first $1000$ ns of training. The following learning parameters have been used for the FA: $\tau_L=62.5$ ns where $C=1$ nF and $R=5 \ \mathrm{k\Omega}$, $A_v=10$ and $R_f=1 \ \mathrm{M\Omega}$. This choice of learning parameters ensures that $\tau_L \gg \tau_N$. Due to the averaging effect of the RC-circuit both sequential and average feeding of the training vector result in similar learning behavior as long as the RC-constant is much larger than the timescale of sequential feeding. Fig. \ref{fig:fig_vivj} c) shows the enlarged version of Fig. \ref{fig:fig_vivj} a). For the sequential feeding, voltage $V_{1,5}$ changes substantially every time $v_1v_5$ switches to -1.

\begin{figure}[!t]
    \setlength\abovecaptionskip{-0.5\baselineskip}
    \centering
    \includegraphics[width=1\linewidth]{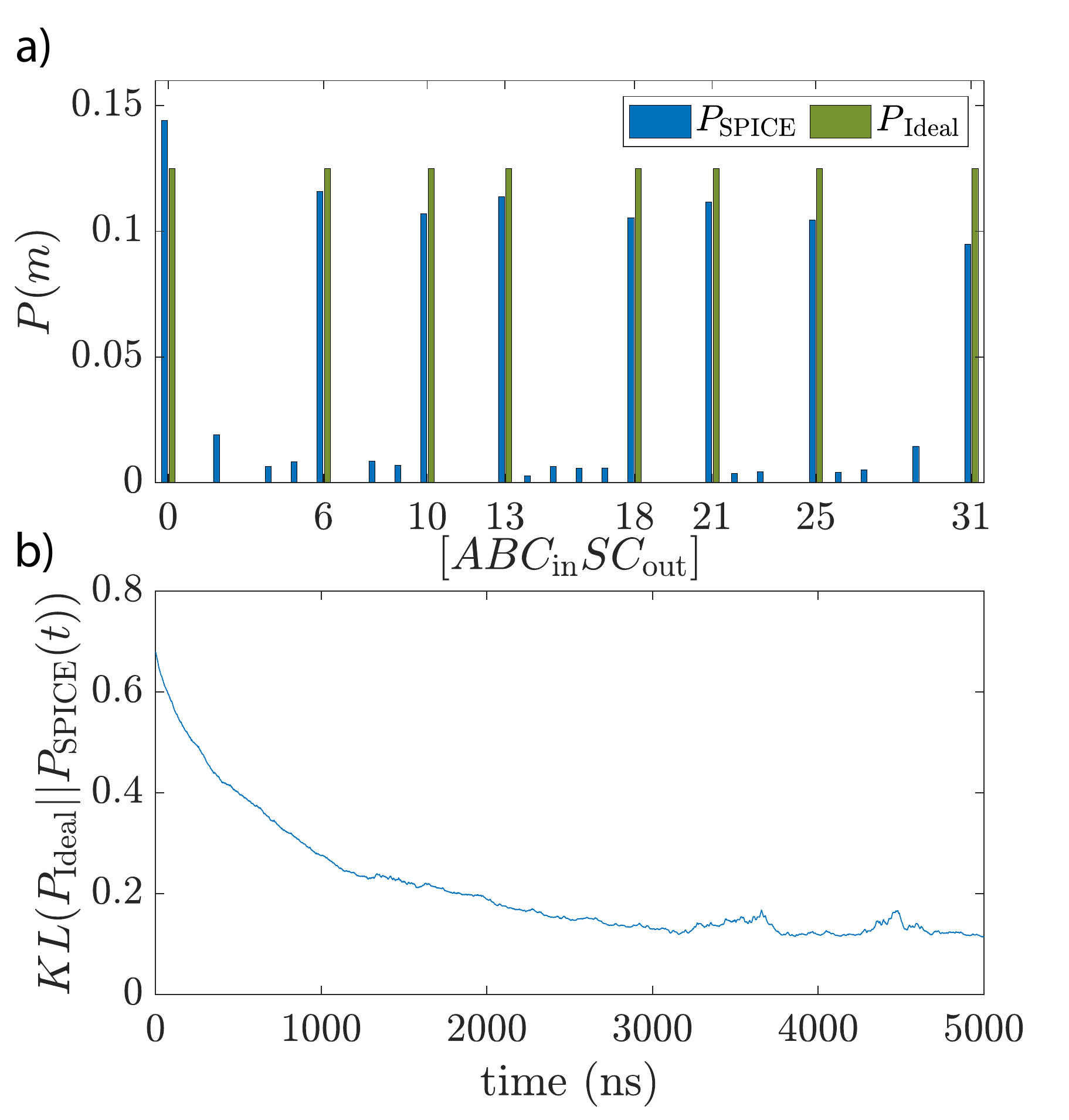} 
    \vspace{0.1in}
    \caption{\textbf{Training of a full adder in SPICE} a) Probability distribution of a trained full adder is compared to the ideal distribution with binary inputs $A$, $B$, $C_\mathrm{in}$ and outputs $S$ and $C_\mathrm{out}$. The training is performed for $5500$ ns. Blue bars are the probability distribution extracted from SPICE simulations by creating a histogram of the configurations of {$m$} over the last $500 \ \mathrm{ns}$ of training. b)  Kullback–Leibler divergence between $P_\mathrm{SPICE}$ obtained by doing a moving average of $500$ ns and the target distribution defined as $KL(P_\mathrm{Ideal} || P_\mathrm{SPICE}(t))=\sum_\mathbf{\mathrm{m}} P_\mathrm{Ideal}(\mathbf{\mathrm{m}}) \log(P_\mathrm{Ideal}(\mathbf{\mathrm{m}})/P_\mathrm{Train}(\mathbf{\mathrm{m}},t))$ . Following parameters have been used in the simulations: $C=1 \ \mathrm{nF}$, $R=5 \ \mathrm{k\Omega}$, $R_F=1 \ \mathrm{M\Omega}$, $A_v=10$, $V_0=50 \ \mathrm{mV}$.}
    \label{fig:fig_FA}
\end{figure}

At the start of training all weight voltages are initialized to 0 V and the probability distribution is uniform. The training is performed for 5500 ns. In Fig. \ref{fig:fig_FA} a) the ideal probability distribution of the FA $P_\mathrm{Ideal}$ is shown together with the normalized histogram $P_\mathrm{SPICE}$ of the sampled BSN configurations taken from the last $500 \ \mathrm{ns}$ of learning  and compared to the ideal distribution $P_\mathrm{Ideal}$. The training vector is fed in as an average. For $P_\mathrm{SPICE}$ the eight trained configurations of table \ref{table: FA} are the dominant peaks. To monitor the training process, the Kullback-Leibner divergence between the trained and the ideal probability distribution $KL(P_\mathrm{Ideal} || P_\mathrm{SPICE}(t))$ is plotted as a function of training time $t$ in Fig. \ref{fig:fig_FA} b) where $P_\mathrm{SPICE}(t)$ is the normalized histogram taken over 500 ns. $P_{\rm SPICE}$ at $t=0$ corresponds to the histogram taken from $t=0$ to $t=500$ ns. During training the KL divergence decreases over time until it reaches a constant value at about 0.1. It has to be noted that after the weight matrix is learned correctly for a fully visible Boltzmann machine, the KL divergence can be reduced further by increasing all weights uniformly by a factor $I_0$ which corresponds to inverse temperature of the Boltzmann machine.\cite{aarts_simulated_1989} Fig. \ref{fig:fig_FA} shows that the probability distribution of a FA can be learned very fast with the proposed autonomous learning circuit. In addition, the learning performance is robust when components of the circuit are subject to variation. In the supplemental material, additional figures of the learning performance are shown when the diameter of the magnet and the resistances of the RC-circuits are subject to variation. The robustness against variations can be explained by the fact that the circuit can learn around variations.  BSNs using LBMs under variations have also been analyzed by Drobitch and Bandyopadhyay\cite{Drobitch_reliability_2019} and Abeed and Bandyopadhyay.\cite{Abeed_Low_2019}\\

\subsection*{Learning image completion}

\begin{figure}[!t]
    \setlength\abovecaptionskip{-0.5\baselineskip}
    \centering
    \includegraphics[width=1\linewidth]{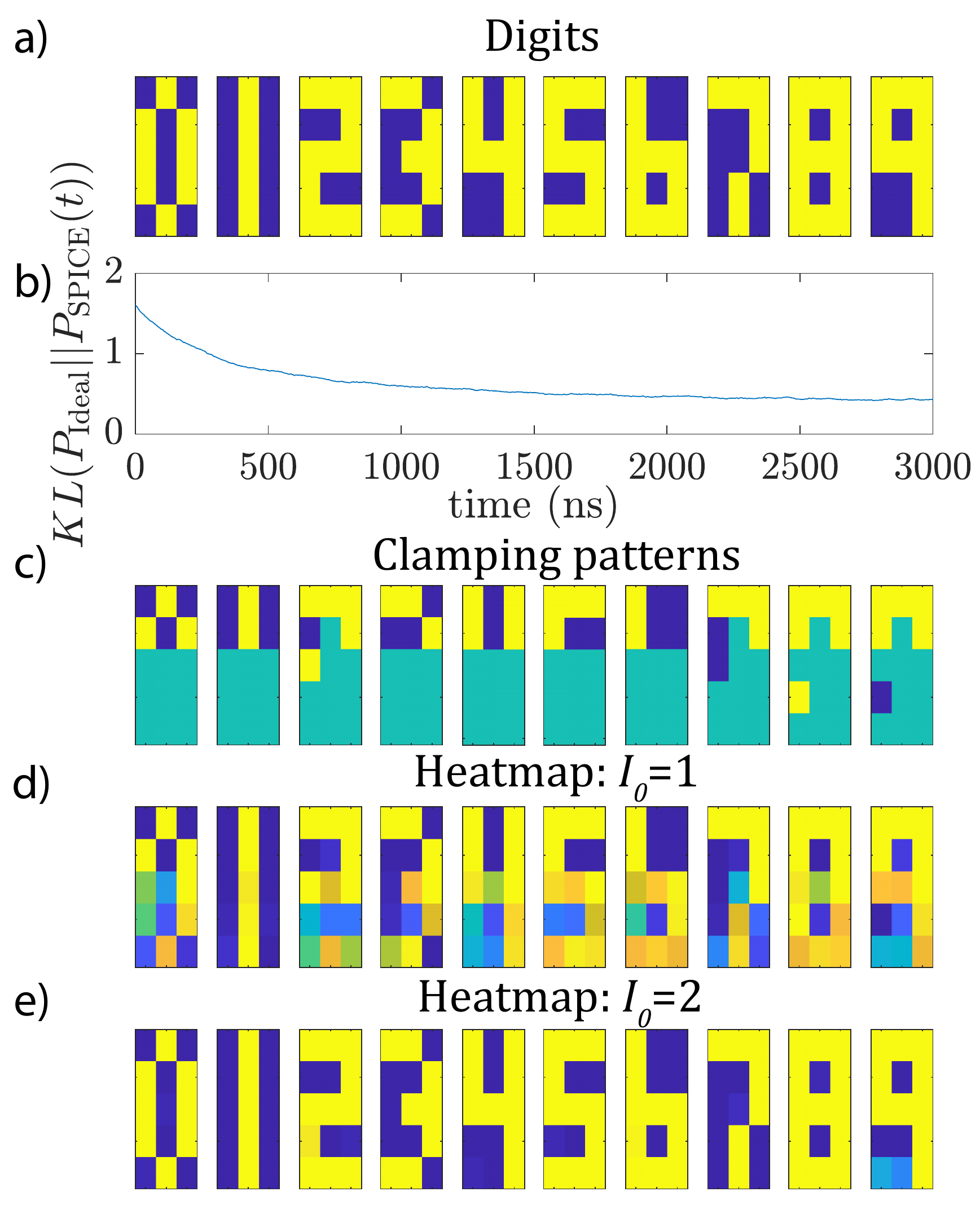} 
    \vspace{0.1in}
    \caption{\textbf{Training and Testing of 5x3 Digit images} a) 5x3 digit images from 0 to 9. b) Kullback Leibner divergence during training for 3000 ns using the autonomous circuit. c)-e) Image completion: For inference, 6 unique pixels are clamped for every digit (as shown in c)). Subfigure d) ande) show the heatmap of BSN outputs during inference for running the circuit for 100 ns for d) $I_0=1$ and e) $I_0=2$.}
    \label{fig:fig4}
\end{figure}

As second example, the circuit is utilized to train 10 5x3 pixel digit images shown in Fig.\ref{fig:fig4} a). Here, 105 reciprocal weights and 15 biases have to be learned. The network is trained for 3000 ns and the bipolar training data is fed in as average of the 10 $v_iv_j$ terms for every digit. The same learning parameters as in the previous section are used here. In \ref{fig:fig4} b) the KL divergence is shown as a function of time between the SPICE histogram and the ideal probability distribution where the ideal distribution has 10 peaks with each peak being 10 \% for each digit. Most of the learning happens in the first 1500 ns of training, however, the KL divergence still reduces slightly during the later parts of learning. After 3000 ns the KL divergence reaches a value of around 0.5.

For inference we replace the capacitor by a voltage source where every voltage is given by the previously learned voltage $V_{ij}$. The circuit is run for 10 instances where every instance has a unique clamping pattern of 6 pixels representing one of the 10 digits. The clamped inputs are shown in Fig. \ref{fig:fig4} c). The input of a clamped BSN is set to $\pm V_{DD}/2$. Each instance is run for 100 ns and the outputs of the BSNs are monitored. The BSNs fluctuate between the configurations given by the learned probability distribution. In Fig. \ref{fig:fig4} d) the heat map of the output of the BSNs is shown. For every digit the most likely configuration is given by the trained digit image. To illustrate this point, the amount of BSN fluctuations is reduced by increasing the learned weight voltages by a factor of $I_0=2$. The circuit is again run in inference mode for 100 ns with the same clamping patterns. In Fig. \ref{fig:fig4} e) the heatmap is shown. The circuit locks in into the learned digit configuration. This shows that in inference mode the circuit can be utilized for image completion.
 
\section{Discussion}
In this paper we have presented a framework for mapping a continuous version of Boltzmann machine learning rule (Eq. \eqref{eq:learning_rule}) to a clockless autonomous circuit. We have shown full SPICE simulations to demonstrate the feasibility of this circuit running without any digital component with the learning parameters set by circuit parameters. Due to the fast BSN operation, samples are drawn at subnanosecond speeds leading to fast learning, as such the learning speed should be at least multiple orders of magnitudes faster compared to other computing platforms. \cite{adachi_application_2015,korenkevych_benchmarking_2016,terenin_gpu-accelerated_2019} The advantage of this autonomous architecture is that it produces random numbers naturally and does not rely on pseudo random number generators like linear-feedback shift register (LFSRs) (which are for example used in Bojnordi et al.\cite{Bojnordi_memristive_2016}). These LFSRs have overhead and are not as compact and efficient as the hardware BSN used in this paper. As shown by Borders et al.\cite{borders_integer_2019}, typical LFSRs need about 10x more energy per flip and more than 100x more area than an MTJ-based BSN. Another advantage of this approach is that the interfacing with digital hardware only needs to be performed after the learning has been completed. Hence, no expensive analog-to-digital conversion has to be performed during learning. We believe this approach could be extended to other energy based machine learning algorithms like equilibrium propagation introduced by Scellier and Bengio\cite{scellier_equilibrium_2017} to design autonomous circuits. Such standalone learning devices could be particularly of interest in the context of mobile and edge computing.

\acknowledgments
This work was supported in part by ASCENT, one of six centers in JUMP, a Semiconductor Research Corporation (SRC) program sponsored by DARPA. KYC gratefully acknowledges support from Center for Science of Information (CSoI), an NSF Science and Technology Center, under grant CCF-0939370.

\balance\bibliography{library}
 \end{document}